\documentclass{ragtime}
\title[Growth of orbital resonances around a black hole surrounded by matter]%
      {Growth of orbital resonances around a black hole surrounded by matter}

\author[M. Straten\'{y} and G. Lukes-Gerakopoulos]
       {Michal Straten\'{y}\at[]{1,2,a} and Georgios Lukes-Gerakopoulos\at[]{1,b}\\
        \ins{1}Astronomical Institute of the Academy of Sciences of the Czech Republic,\splitins[1]
        Bo\v{c}n\'{i} II 1401/1a, CZ-141 31 Prague, Czech Republic\\
        \ins{2}Institute of Theoretical Physics, Faculty of Mathematics and Physics,\splitins[1]
        Charles University, CZ-180 00 Prague, Czech Republic\\
        \ins{a}\Email{strateny.m@gmail.com} \ins{b}\Email{gglukes@gmail.com}}

\usepackage[scr=boondox]{mathalfa}
\usepackage{subfig}
\usepackage{float}
\usepackage[normalem]{ulem}
\usepackage{color}
\definecolor{darkgreen}{rgb}{0.0, 0.65, 0.31}

\begin{document}

\begin{abstract}
    This work studies the dynamics of geodesic motion within a curved spacetime around a Schwarzschild black hole, perturbed by a gravitational field of a far axisymmetric distribution of mass enclosing the system. This spacetime can serve as a versatile model for a diverse range of astrophysical scenarios and, in particular, for extreme mass ratio inspirals as in our work. We show that the system is non-integrable by employing Poincaré surface of section and rotation numbers. By utilising the rotation numbers, the widths of resonances are calculated, which are then used in establishing the relation between the underlying perturbation parameter driving the system from integrability and the quadrupole parameter characterising the perturbed metric. This relation allows us to estimate the phase shift caused by the resonance during an inspiral.
\end{abstract}


\section{Introduction}\label{sec: Introduction}

Even if the Kerr black hole solution of the Einstein field equation is a vacuum one \citep{Kerr1963}, astrophysical black holes are surrounded by matter distributions. Therefore, a curved spacetime describing a Schwarzschild black hole perturbed by the gravitational field of a far axisymmetric distribution is of interest since it can serve as a model for a diverse range of physical scenarios. For instance, massive black holes located in the centres of galaxies are often surrounded by dense nuclear star clusters and other molecular and dust structures~\citep{Genzel2010}. Furthermore, alternative sources of external matter can come from more exotic sources, such as dark matter or scalar fields \citep{Hannuksela2020,Ferreira2017}.

In the close vicinity of these primary massive black holes, we expect that stellar compact objects are trapped by the gravitational field of the primaries. While such a secondary object orbits the primary one, it loses energy and angular momentum in the form of gravitational waves. This loss leads the secondary to inspiral towards the primary creating an extreme mass ratio inspiral (EMRI). These gravitational waves peak in the mHz frequency band and are expected to be observed by the next generation of gravitational-wave observatories \citep{Berry2019,AmaroSeoane2017, Polcar2022}.

The perturbation of a Schwarzschild black hole by the surrounding matter can have observational implications on the evolution of an EMRI. In particular, a key aspect of geodesic motion in the Schwarzschild spacetime is its integrability. However, by perturbing the spacetime, the system is no longer entirely symmetrical, which results in the loss of integrals of motion. The insufficient number of integrals of motion leads to non-integrability, which allows chaotic behaviour to emerge.

We focus mainly on resonances, parts of the phase space where two or more characteristic frequencies of the system match in integer ratios. These regions are key parts of the study of chaos because there chaotic behaviour emerges. There are studies \citep[see, e.g.,][ and references therein]{Lukes-Gerakopoulos2022} showing that such regions can have an observational impact on the gravitational waves emitted during an EMRI, since a resonance crossing is expected to cause a dephasing of the gravitational waveforms. Hence, investigating the strength and the growth of the resonances is important for the preparation of the gravitational waveforms needed to detect the signal from an EMRI \citep{Babak2017}.

The article is organised as follows: in section~\ref{sec: Geodesic motion}, we introduce the properties of the studied spacetime, some cardinal theoretical elements of non-integrable systems, along with some tools to study these systems. Section~\ref {sec: Numerical Examples} showcases our numerical results obtained from the computation of geodesic motion within the studied curved spacetime. Section~\ref{sec: Conclusion} discusses our main findings. Note that geometric units are employed throughout the article $G=c=1$. Greek letters denote the indices corresponding to spacetime. The metric signature is chosen as $\left( -+++ \right)$.

\section{Geodesic motion and chaos}\label{sec: Geodesic motion}

This work studies geodesic motion in curved spacetime introduced by~\cite{Polcar2022}, who outlined its derivation. This metric expresses a spacetime around a non-spinning black hole of mass $M$ encircled by a rotating gravitating ring with mass $\mathscr{M}_r$ at a radius $r_r \gg M$~much larger than the black hole horizon. The gravitational field near the black hole is influenced by the tidal effects caused by the presence of the ring. The resulting spacetime is static and axisymmetric, while the multipole structure of the ring is truncated to the leading quadrupolar order. In Schwarzschild-like coordinates $(t,r,\theta,\phi)$, the resulting metric is given by the line element:

\begin{equation}\label{eq: Metric}
\begin{split}
{ds}^2_{r \ll r_r} = 
& -\left(1-\frac{2M}{r}\right)\left(1+2\nu_\mathscr{Q}\right)dt^2 + \frac{1+2\chi_\mathscr{Q}-2\nu_\mathscr{Q}}{1-2M/r}dr^2 \\
& + (1-2\nu_\mathscr{Q})r^2\left[(1+2\chi_\mathscr{Q})d\theta^2+\sin^2\theta d\phi^2\right] ,
\end{split}
\end{equation}

\begin{subequations}\label{eq: Metric parts}
\begin{alignat}{4}
\nu_\mathscr{Q} &\equiv \frac{\mathscr{Q}}{4}\left[r(2M-r)\sin^2\theta+2(M-r)^2\cos^2\theta-6M^2\right] , \\
\chi_\mathscr{Q} &\equiv \mathscr{Q} M(M-r)\sin^2\theta ,
\end{alignat}
\end{subequations}
with $\mathscr{Q} \equiv \mathscr{M}_r/r_r^3$ representing the quadrupole perturbation parameter. It should be noted that metric~\eqref{eq: Metric} is valid only for $r \ll r_r$ and assumes non-compact rings, i.e. $\mathscr{M}_r \ll r_r$. Especially, it neglects all terms starting from $\mathscr{O} \left( r_r^{-4} \right)$ and $\mathscr{O} \left( \mathscr{M}_r^{2} \right)$.

The Hamiltonian function governing the geodesic evolution of the system is:
\begin{equation}\label{eq: Hamiltonian}
\mathcal{H}= \frac{1}{2}g^{\mu \nu}p_\mu p_\nu =-\frac{1}{2}m^2,
\end{equation}
where $m$ is the mass of the secondary body, while one of Hamilton's equations reads
\begin{equation}\label{eq: Integrals of motion from Hamiltonian}
\frac{d p_\kappa}{d\tau} = - \frac{\partial \mathcal{H}}{\partial q^{\kappa}} = -\frac{1}{2}g_{\mu \nu, \kappa}p^\mu p^\nu ,
\end{equation}
where $\tau$ is the proper time and we utilised that the canonical momenta are independent of the generalised coordinates; therefore, only the metric tensor will contribute to the derivative with respect to generalised coordinates. This implies that if the metric tensor is independent of a particular generalised coordinate, the conjugate momentum associated with that coordinate is conserved and considered an integral of motion.
This observation corresponds with Noether's theorem, which states that symmetries of a system give rise to the existence of integrals of motion. Furthermore, it implies the presence of a Killing vector field in the differential geometry. 

The initial system exhibits four degrees of freedom; however, if we examine the form of the metric~\eqref{eq: Metric}, we see that it does not depend on the time variable $t$ and the azimuthal angle $\phi$. The conjugate momentum associated with the time coordinate is denoted as $p_t \equiv -E$ and it has the meaning of the \textit{total energy}, whereas the conjugate momentum related to the azimuthal angle is denoted as $p_\phi \equiv L$ and it has the meaning of a \textit{angular momentum} \citep[see, e.g.][]{Misner2017}. Both quantities are referenced with respect to an observer standing at infinity. These symmetries allow us to reduce our system to a two degrees of freedom problem solely described by coordinates $r$ and $\theta$. The reduced Hamiltonian of this system takes the following form:
\begin{equation}\label{eq: Reduced Hamiltonian}
\mathcal{H} = \frac{1}{2} \left( \frac{\left( p_r \right)^2}{g_{rr}}  + \frac{\left( p_\theta \right)^2}{g_{\theta \theta}} + \frac{E^2}{g_{tt}} + \frac{L^2}{g_{\phi \phi}} \right).
\end{equation}
Because of the Liouville-Arnold theorem \citep{Arnold1989}, in the reduced Hamiltonian system of two degrees of freedom, the bounded motion takes place on a two-dimensional invariant torus with two fundamental frequencies. We can distinguish the nature of the motion by the ratio of these fundamental frequencies $\omega = \omega^1 / \omega^2$. If $\omega$ is irrational, the motion on the torus is quasiperiodic. In this case, the quasiperiodic orbit densely covers the torus over an infinite amount of time and does not return to its initial state from where it started within a finite period. If $\omega$ is rational, the torus is called resonant, and the motion is periodic. For a detailed analysis of the resonance condition, we refer to \citet{Lukes-Gerakopoulos2022}.

Let us consider an integrable system that undergoes perturbation, causing it to lose its integrability, enabling chaos to occur. The transition from integrable two degrees of freedom systems to non-integrable ones is governed by two fundamental theorems: the Kolmogorov-Arnold-Moser theorem (abbreviated as the KAM theorem) and the Poincaré-Birkhoff theorem (for detailed references see~\cite{Iro2016}). According to the KAM theorem, the non-resonant torus survives for small perturbations. These tori are called KAM tori. Furthermore, Poincaré-Birkhoff theorem states that in spaces where was a resonant torus, an even number of periodic trajectories survive; half of them are stable and half unstable.

\begin{figure}[t]
\centering
\subfloat[$\omega^{1}/\omega^{2} = 4/5$]{
  \includegraphics[width=0.48\linewidth,height=0.20\linewidth]{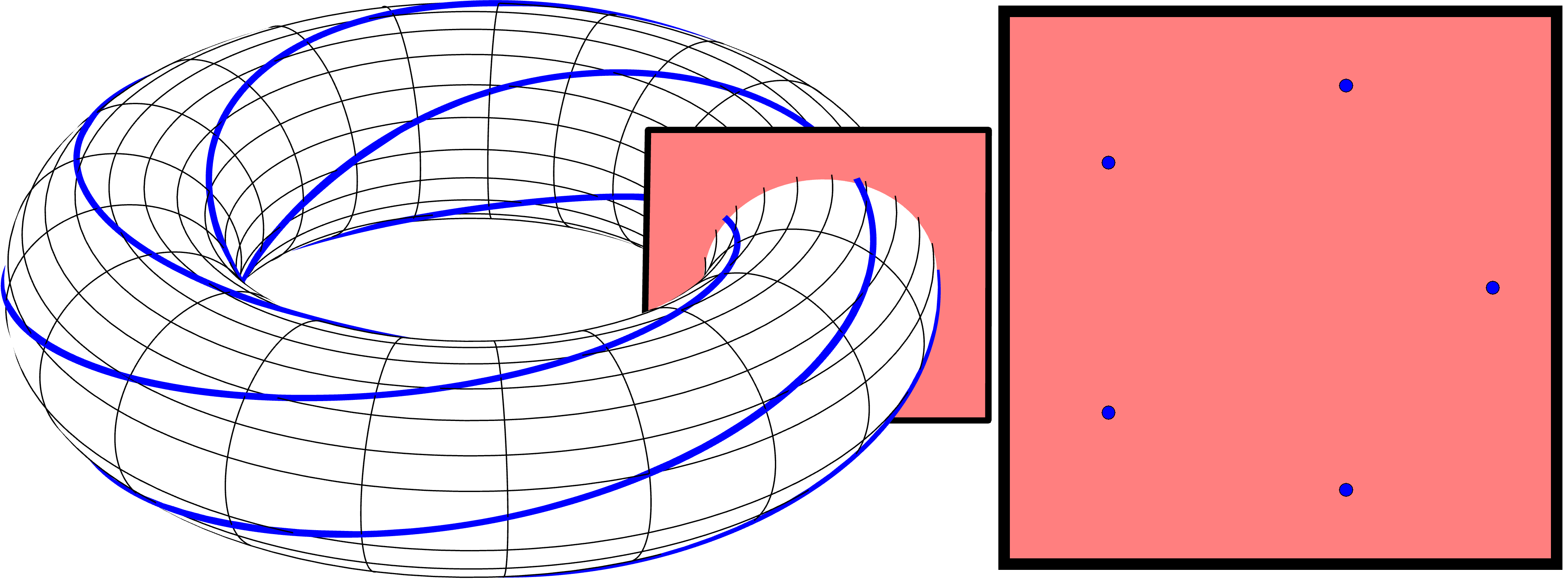}
}
\subfloat[$\omega^{1}/\omega^{2} \approx 0.8001$]{
  \includegraphics[width=0.48\linewidth,height=0.20\linewidth]{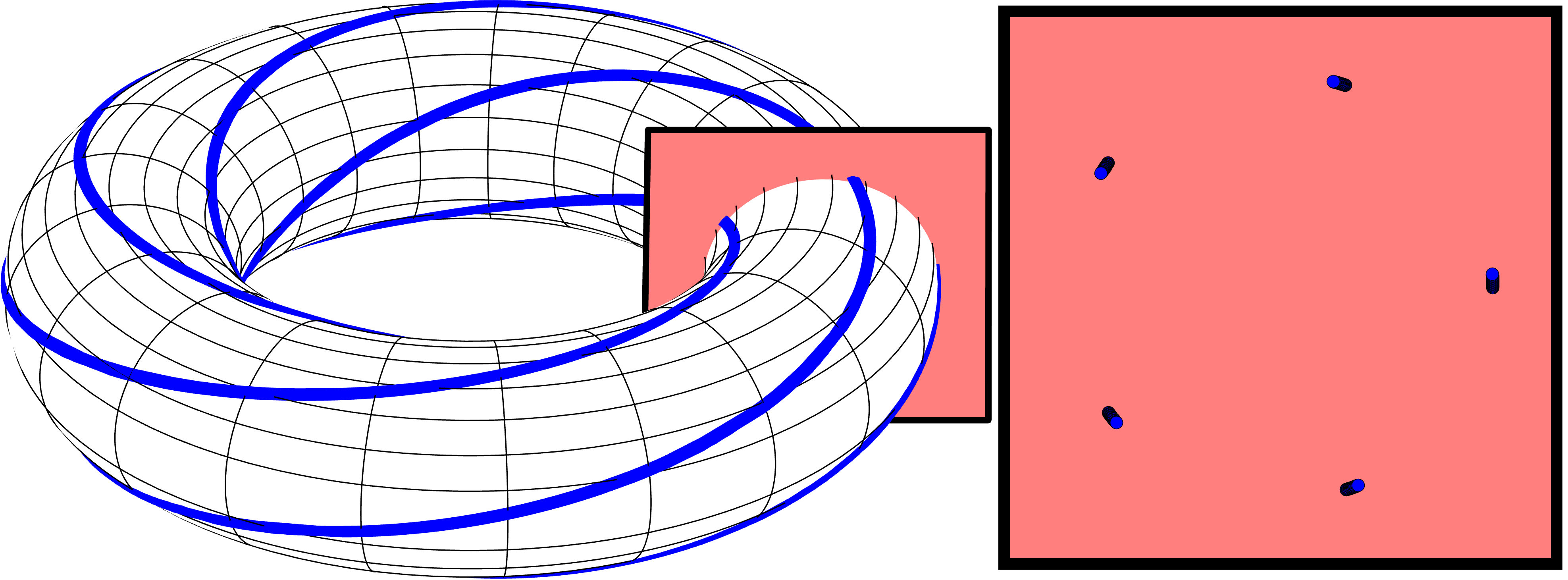}
}
\hspace{0mm}
\subfloat[$\omega^{1}/\omega^{2} \approx 0.801$]{
  \includegraphics[width=0.48\linewidth,height=0.20\linewidth]{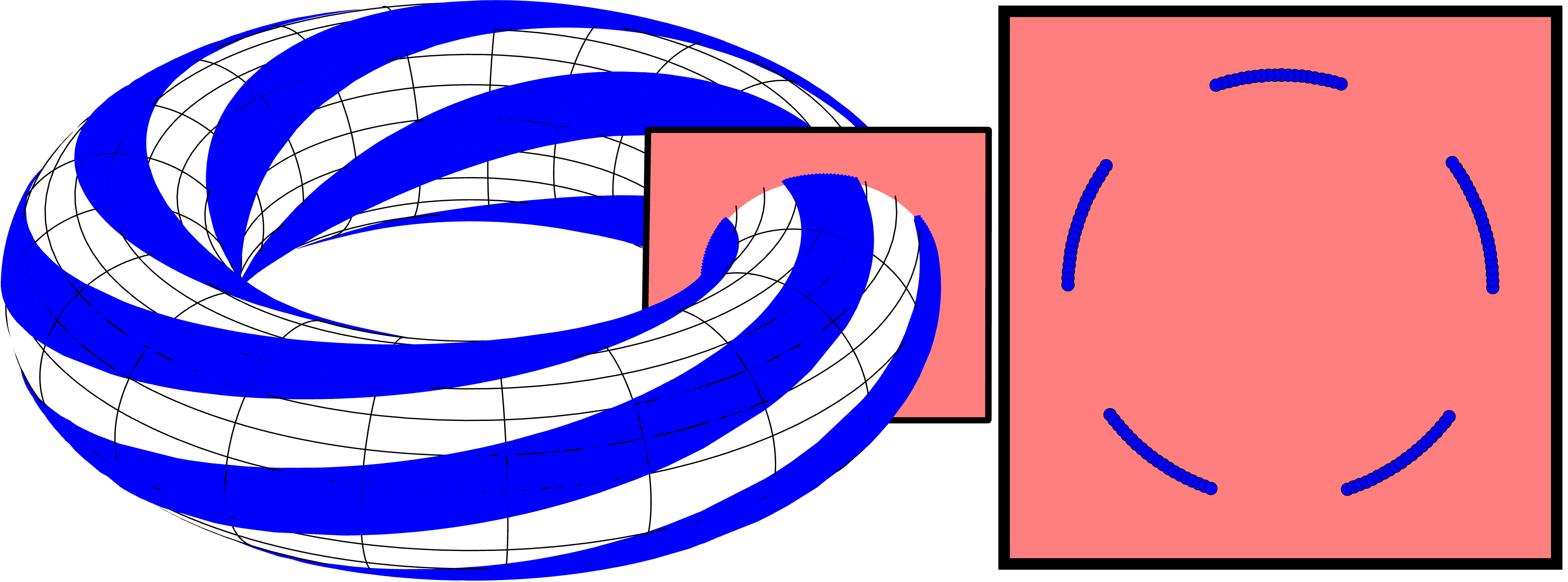}
}
\subfloat[$\omega^{1}/\omega^{2} \approx 0.81$]{
  \includegraphics[width=0.48\linewidth,height=0.20\linewidth]{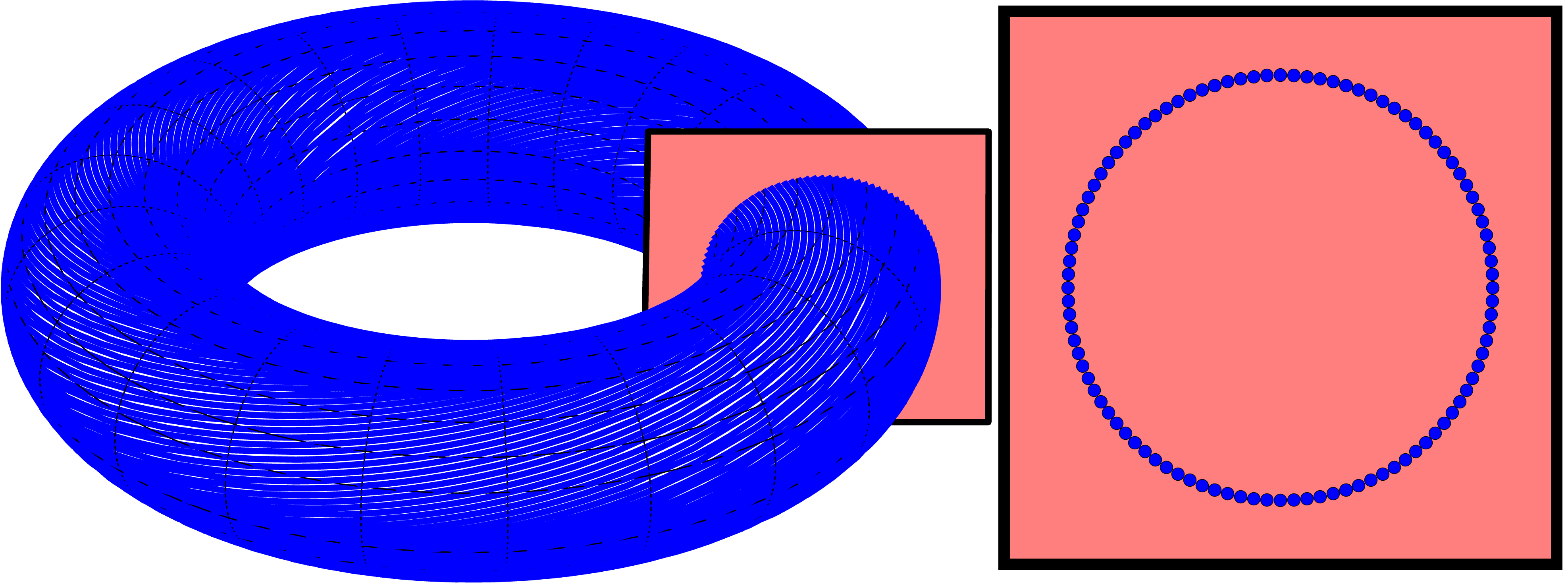}
}
\caption{An illustration of a resonant (a) torus and three quasiperiodic ones (b,c,d) in a two degree of freedom system. The trajectories on the tori are depicted in blue, while the Poincaré surfaces of section are in red. Each trajectory undergoes calculations until the motion intersects the Poincaré surface of section for a total of 100 occurrences. The resonant trajectory, located in the upper left plot, exhibits a resonant ratio expressed as $\omega^{1}/\omega^{2} = 4/5$, representing the fundamental frequency of the small circle over the large circle. In contrast, the remaining trajectories represent non-resonant trajectories, characterised by ratios that deviate from the resonant ratio and are of an irrational nature. Over time, the non-resonant trajectories asymptotically trace out the torus cross-section.}
\label{fig: Torus with Poincare sections}
\end{figure}

\begin{figure}[t]
\begin{center}
\includegraphics[width=0.92\linewidth]{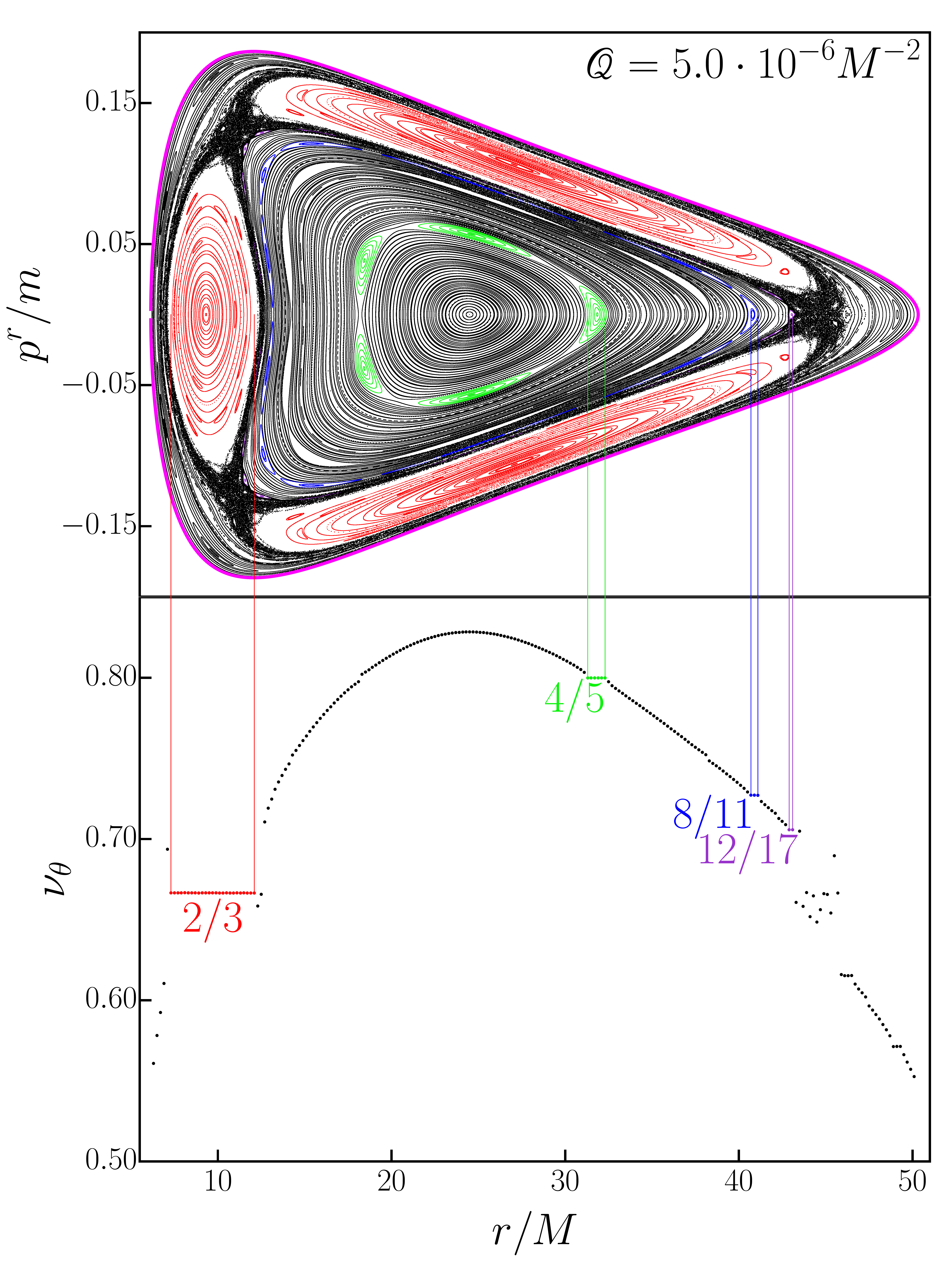}
\end{center}
\caption{The top figure displays a Poincaré surface of section, while the bottom figure shows the corresponding rotation curve computed along the ${p^r/m=0}$ line. The dominant resonances are prominently marked in both figures, along with their ratios of fundamental frequencies. The parameters taken are ${L=4.0m M}$, ${E=0.98 m}$, ${\theta \left[ 0 \right]= \pi/2}$ and ${r \left[ 0 \right] \in \left( 6.298M; 50.098M \right)}$ with step size $0.2M$.}
\label{fig: Poincare section}
\end{figure}

To study the dynamics of a two degree of freedom Hamiltonian system, the symplecticity of the system allows us to reduce the 4-dimensional phase space to a two dimensional section in it, known as \textit{Poincaré surface of section}. In order to achieve that, we have to find a section through the tori foliation, where the Hamiltonian flow is transverse. A schematic of such section is shown in Fig.~\ref{fig: Torus with Poincare sections}, where it is shown how the motion on torii is mapped on a surface of section. In particular, Fig.~\ref{fig: Torus with Poincare sections}(a) shows how a periodic orbit on a resonance torus is mapped on a surface of section, while Figs.~\ref{fig: Torus with Poincare sections}(a)-(c) show how quasiperiodic orbits on non-resonance tori are mapped on a surface of section. To computationally create a Poincaré section, one must integrate the equations of motion and identify the constant of the motion that remains fixed, as well as the section condition that reflects the symmetries of the system. Only two remaining phase-space coordinates are recorded once the trajectory passes through this section of surface. After a sufficient number of crossings have been recorded, the results can be plotted.

The system we study exhibits reflection symmetry along the equatorial plane; thus, in order to ensure that the chosen section intersects the Hamiltonian flow perpendicularly, we set the Poincaré surface of section as the equatorial plane (i.e.~$\theta \equiv \pi /2$). Additionally, we only consider points intersecting the surface of section from a specific direction, which we choose to be $u^\theta >0 $. Therefore, we are left with the two remaining phase coordinates, $r$ and $p^r$, which we record (Fig.~\ref{fig: Poincare section}). The accessible for the motion region on the Poincaré surface of section is determined by Eq.~\eqref{eq: Reduced Hamiltonian} \citep[see, e.g.,][]{Polcar2019}. We represent the boundary curve of accessible region as a thick magenta curve in Fig.~\ref{fig: Poincare section}.

Resonances are parts of the phase space where the non-integrable behaviour emerges. Therefore, having methods to identify these zones effectively would be advantageous. In systems with two degrees of freedom, we might take advantage of the Poincaré surface of section to evaluate the rotation number $\nu_\vartheta$ \citep[for further details, see e.g.][]{Lukes-Gerakopoulos2022,Voglis1999}. Assuming we have already produced a Poincaré surface of section of the system, we must first distinguish the centre of the main island of stability, represented by a fixed point $\mathbf{x}_s$ on the Poincaré section, around which most invariant curves are nested. Next, we evaluate rotation angle, defined as the angle between two vectors originating from~$\mathbf{x}_s$ and pointing towards two consecutive points on the Poincaré surface of section:
\begin{equation}\label{eq: Rotation angle}
\vartheta_i := ang \bigl[ \left( \mathbf{x}_{i+1} - \mathbf{x}_c \right) ; \left( \mathbf{x}_i - \mathbf{x}_c \right) \bigr].
\end{equation}

The angle value spectrum has to be restricted to a proper interval so there is no discontinuity in the spectrum \citep{Voglis1999}. The \textit{rotation number} is then obtained as the average of these rotation angles as follows:
\begin{equation}\label{eq: Rotation number}
\nu_\vartheta = \lim_{N\to\infty} \frac{1}{2\pi N}\sum^N_{i=1}\vartheta_i.
\end{equation}
In the limit, $N \rightarrow \infty$, the calculated rotation number~\eqref{eq: Rotation number} corresponds to the ratio of two fundamental frequencies $\omega^1 / \omega^2$. For finite $N$, the inaccuracy of calculations is approximately equal to \citep{Voglis1999}:
\begin{equation}\label{eq: Inaccuracy of rotation numbers}
\delta_\vartheta = \frac{\Delta}{N}, \text{ where } 0<\Delta<1 .
\end{equation}
Hence, in order to obtain more precise values of rotation numbers, it is necessary to record more rotation angles. Typically, we have calculated the trajectory for our numerical results until 35000 rotation angles have been recorded.

The plot of rotation numbers $\nu_\vartheta$ as a function of distance from the main island's centre is referred to as a \textit{rotation curve}.
The rotation curve is strictly monotonic for integrable systems as one advances away from the centre $\mathbf{x}_s$. In contrast, for perturbed non-integrable systems, the curve maintains qualitative similarity to the unperturbed one, except in the vicinity of the resonances where it exhibits significant changes. The resonance curve starts to fluctuate randomly in chaotic layers at the resonance (Fig.~\ref{fig: Poincare section}). Moreover, within the Birhoff chain, stable regions known as islands of stability appear as plateaus with constant values in the rotation curve (Fig.~\ref{fig: Poincare section}).

The \textit{width of a resonance} is a useful measure as it can be related to the perturbation parameter $\epsilon$ driving the system away from integrability.
It can be shown that the relation between the width of the resonance and the perturbation parameter of two degrees of freedom system, as provided by~\cite{Lukes-Gerakopoulos2022} \citep[further details are given by][]{Arnold2007} can be expressed as:
\begin{equation}\label{eq: Width of resonance}
w:= 4\sqrt{\frac{\alpha}{\beta}}\sqrt{\epsilon},
\end{equation}
where $w$ denotes the width of the resonance and $\alpha,\beta$ are positive parameters.
By taking the logarithm of the equation~\eqref{eq: Width of resonance}, we obtain:
\begin{equation}\label{eq: Logarithmic width}
\log{w \left( \mathscr{Q} \right)}= \frac{1}{2}\log{\epsilon \left( \mathscr{Q} \right)}+\log{4\sqrt{\frac{\alpha}{\beta}}} ,
\end{equation}
where we emphasise that on the right-hand side of the equation, only $\epsilon$ is dependent on $\mathscr{Q}$ as $\alpha$ and $\beta$ are positive parameters. Therefore, by plotting the width of the resonance with respect to the quadrupole perturbation parameter on a logarithmic scale and performing the linear regression, we are able to quantify the relation between $\epsilon$ and $\mathscr{Q}$.
The aforementioned linear regression takes the following form:
\begin{equation}\label{eq: Linear regresion}
\log{\frac{w \left( \mathscr{Q} \right)}{M}} = A \cdot \log{\mathscr{Q}M^2} + B .
\end{equation}
Thus, by comparing the equations~\eqref{eq: Logarithmic width} and~\eqref{eq: Linear regresion}, we obtain the following power law expression for the perturbation parameter:

\begin{equation}
\epsilon = (\mathscr{Q}M^2)^{2 A} .
\end{equation}

\section{Numerical Examples}\label{sec: Numerical Examples}

In this section, we aim to employ the theoretical concepts introduced in the previous section to examine the orbital resonances in the spacetime described by the metric~\eqref{eq: Metric}. For our numerical calculations, we kept fixed ${E=0.98m}$, ${L=4.0mM}$, and the initial conditions were chosen along the $p^r/m=0$ line in the Poincaré surface of section.

\begin{figure}[t]
\begin{center}
\includegraphics[width=\linewidth]{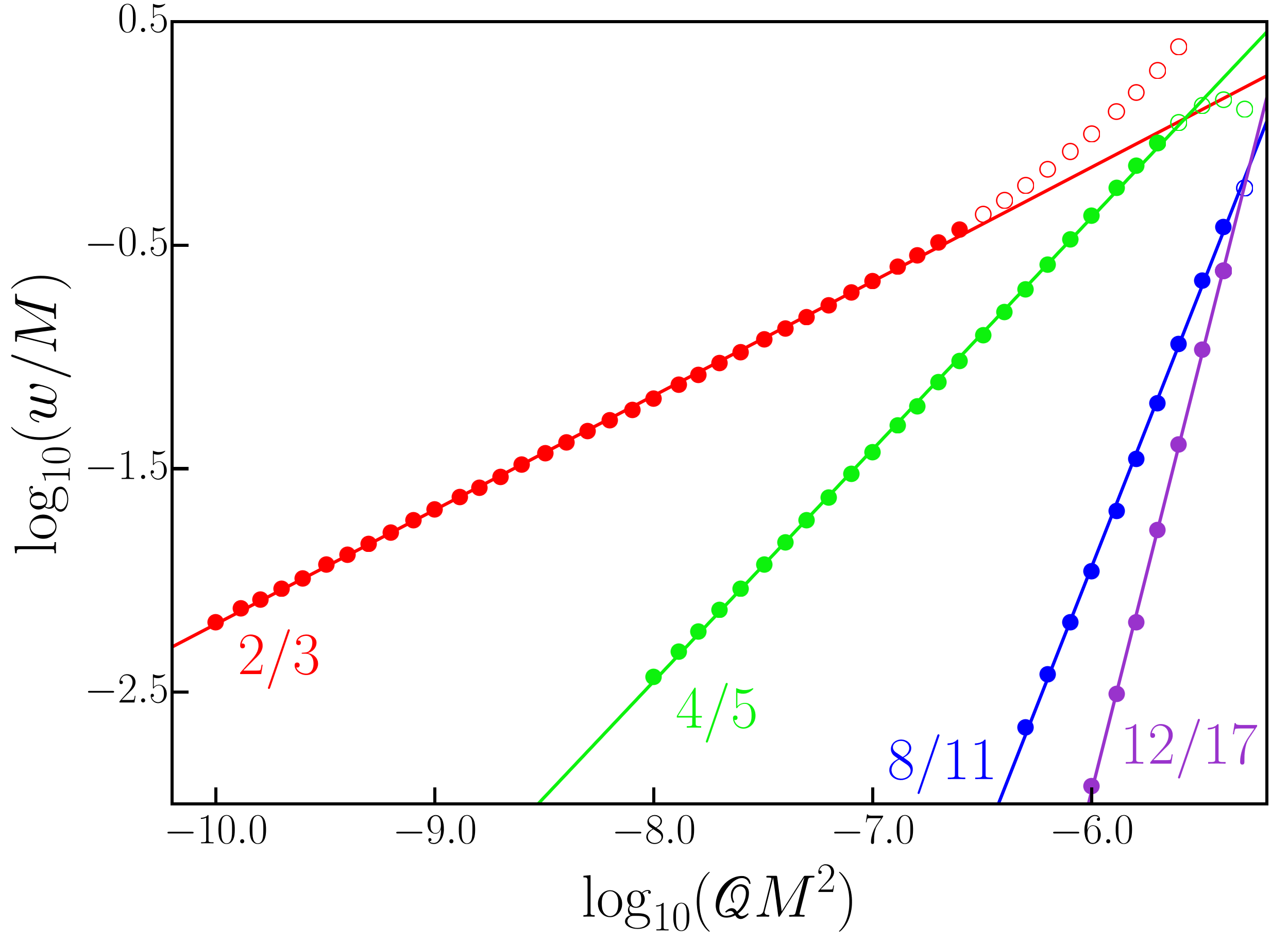}
\end{center}
\caption{Logarithmic plot of widths of resonances with respect to the quadrupole perturbation parameter. The figure omits the error bars as their size is smaller than the plot symbols (the relative error was maintained below~$1\%$).}
\label{fig: All resonances}
\end{figure}

\begin{table}[t]
    \centering
    \begin{tabular}{cccc}
\hline 
\rule{0pt}{10pt}  \textbf{Resonance} & \textbf{Parameter A} & \textbf{Parameter B} & \textbf{Power law} \\
\hline 
\rule{0pt}{10pt} 2/3   & $0.511 \pm 0.002$ & $2.91 \pm 0.01$  & $\epsilon = \mathscr{Q}M^2$ \\
\hline 
\rule{0pt}{10pt} 4/5   & $1.038 \pm 0.004$ & $5.85 \pm 0.03$  & $\epsilon = \mathscr{Q}^2M^4$ \\
\hline
\rule{0pt}{10pt} 8/11  & $2.491 \pm 0.027$ & $13.01 \pm 0.16$ & $\epsilon = \mathscr{Q}^5M^{10}$ \\
\hline
\rule{0pt}{10pt} 12/17 & $3.880 \pm 0.040$ & $20.34 \pm 0.23$ & $\epsilon = \mathscr{Q}^8M^{16}$ \\
\hline
\end{tabular}
\caption{Results of resonance growth with respect to quadrupole perturbation parameter.}
    \label{Tb: Summary of parameters}
\end{table}

Fig.~\ref{fig: Poincare section} showcases Poincaré section along with the corresponding rotation curve. In the particular figure, it is evident that the presence of a chaotic layer is not negligible; therefore, one must be cautious in using the widths of resonances in such cases. Namely, prominent chaotic layers, in general, tend to distort the width of resonance, since we depart from the pendulum approximation used in establishing the relation~\ref{eq: Width of resonance}; thus, these data points cannot be used in the linear regression analysis. Nonetheless, the Fig.~\ref{fig: Poincare section} serves as a valuable example wherein all four resonances are discernible.

For several values of the quadrupole perturbation parameter, the widths of the four most prominent resonances were recorded. The results can be seen in Fig.~\ref{fig: All resonances} where a linear regression is applied. During the linear regression, certain data points were omitted (marked in a particular figure with empty circles). This exclusion is due to those points corresponding to the quadrupole perturbation parameter, where a prominent chaotic layer surrounds the particular resonance. Table~\ref{Tb: Summary of parameters} provides a summary of the power laws for the studied resonances. We see from the table~\ref{Tb: Summary of parameters} that the relation between the quadrupole perturbation parameter $\mathscr{Q}$ and the perturbation parameter $\epsilon$ is not uniform and differs for each resonance. This is not in agreement with previous studies~\citep{Lukes-Gerakopoulos2022,Zelenka2020,Mukherjee2023}, which indicated one single relation. Therefore, we verified that the validity of such a relation is not necessarily global and can differ from resonance to resonance in the phase space.

\section{Discussion}\label{sec: Conclusion}

Apart from the energy and the $L$, in the Schwarzschild case, there is the total angular momentum constant. The total angular momentum is not constant of motion after the quadrupole perturbation is imposed. The absence of this constant drives the system away from integrability. For small perturbation values, a total angular momentum-like quantity should oscillate around an averaged value \citep{Polcar2022,Kerachian2023}. We speculate that the order $\mathcal{O}(Q^n)$ at which the Poisson bracket of this quantity with the Hamiltonian departs from zero near each resonance should correspond to the value we are finding. However, this speculation is yet to be investigated. What we can deduce from the obtained results is that each resonance appears to be driven from a different order in the perturbation. Moreover, it appears that the value of $n$ in $\mathcal{O}(Q^n)$ grows as the denominator in the resonance ratio increases.

The values of the quadrupole perturbation parameter, like $\mathscr{Q}=10^{-7} M^2$, used in this work were slightly exaggerated. Namely, by taking into account that the quadrupole perturbation parameter is defined as $\mathscr{Q} \equiv \mathscr{M}_r/r_r^3$, the radius of the gravitating ring should be, for instance, at $r_r=100M$ and the mass of the ring should be $\mathscr{M}_r=0.1M$, which implies that the mass of the accreting matter in the near vicinity of the primary black hole corresponds to a significant fraction of its mass. However, the findings remain interesting, even if they do not correspond to probable astrophysical scenarios, for which $\mathscr{Q}$ should be much smaller, since they provide an insight into the influence of the accreting matter on the resonances. With respect to the perturbation parameter and $\mathscr{Q}$ relation, the only astrophysically relevant resonance is $2/3$, rendering the remaining resonances negligible. Thus, it is possible to utilise the linear relation between the perturbation parameter and the quadrupole perturbation parameter in most regions of the phase space.

\ack
M.S. and G.L-G have been supported by the fellowship Lumina Quaeruntur No. LQ100032102 of the Czech Academy of Sciences. We would like to thank Vojt\v{e}ch Witzany for his remarks.

\bibliography{ragsamp}

\end{document}